\documentclass[]{jpconf}
\usepackage{graphicx}
\usepackage{feynmp-auto}
\usepackage{amsmath}
\usepackage{graphicx,epsfig,float,color}
\usepackage{hyperref}
\usepackage{calrsfs}
\usepackage{blindtext}
\usepackage[square,sort,comma,numbers]{natbib}
\usepackage{subfig}
\usepackage[labelfont=bf]{caption}

\newcommand{\pT}{\ensuremath{p_{\text{T}}}\xspace}

\newcommand{\ggf}{$gg$F\xspace}

\begin{document}
\title{Production of the Madala boson in association with top quarks}

\author{Stefan von Buddenbrock}
\address{School of Physics, University of the Witwatersrand, Johannesburg 2050, South Africa.}
\ead{stef.von.b@cern.ch}

\begin{abstract}
The Madala hypothesis is the prediction of a new heavy scalar, the Madala boson, that has had previous success in explaining several anomalies in LHC Run 1 and 2 data.
In the literature, the Madala boson has so far primarily been discussed in the context of its dominant production mode, gluon fusion.
However, it can be shown that a study of its production in association with top quarks can provide us with crucial information about the model, as well as explain the enhancement of top associates Higgs production that has been observed in the data -- most notably in leptonic channels.
For this study, Monte Carlo events have been produced and passed through a detector simulation.
These events are then run through an event selection designed by a CMS search for a single top quark in association with a Higgs boson.
A fit is made to the CMS data, yielding a parameter constraint on the Madala hypothesis.
With the Madala hypothesis prediction, an effective signal strength is calculated and compared with the observed values.
\end{abstract}

\section*{Introduction}

The search for physics beyond the Standard Model (BSM) has gained considerable interest since the discovery of the Standard Model (SM) Higgs boson, $h$~\cite{Aad:2012tfa,Chatrchyan:2012xdj}.
In the years since, a plethora of models extending the SM have been proposed with the potential of being discovered experimentally at the Large Hadron Collider (LHC) -- for a recent review, see Ref.~\cite{Csaki:2018muy}.
One such model that has previously been discussed in the literature is known as the \textit{Madala hypothesis}.
The key postulate of this model is the hypothetical existence of a heavy scalar $H$ -- the \textit{Madala boson} -- which interacts strongly with the SM Higgs boson and an additional Higgs-like scalar singlet $S$ -- which can act as a portal to some arbitrary BSM physics~\cite{vonBuddenbrock:2016rmr}.
Applying the hypothesis to several Run 1 experimental results also placed constraints on the mass of $H$, with a best fit point at $m_H=272^{+12}_{-9}$~GeV~\cite{vonBuddenbrock:2015ema}.
The mass of $S$ has not yet been constrained at the present time of writing, but it is considered in the range $m_S\in[130,200]$~GeV.
This is due to the fact that a rich combination of the production of multiple leptons can be explored in this mass range~\cite{vonBuddenbrock:2016rmr}, due to the fact that the the $S$ will decay dominantly to pairs of massive gauge bosons (as opposed to $b$-quarks, which are the dominant decays for Higgs-like scalars with lower masses).

So far, the majority of studies done on the Madala hypothesis have looked dominantly at the gluon fusion (\ggf) production mechanism of $H$.
While \ggf is indeed assumed to have a large production cross section, it should be noted that the same theoretical vertices required for \ggf require that $H$ can also be produced in association with top quarks ($ttH$) with a non-negligible cross section.
This is a tantalising prospect, since both Run 1 and Run 2 searches for top associated Higgs production ($tth$) have shown significant excesses, particularly in leptonic channels.
When measuring the signal strength of $tth$ using $\mu_{tth}=\sigma_{tth}^\text{obs}/\sigma_{tth}^\text{SM}$, it can be shown that a combination of the experimental search results in leptonic channels yields a value of $\mu_{tth}=1.92\pm0.38$~\cite{vonBuddenbrock:2017bqf}.
This can be quantified as around a $2.4\sigma$ deviation from the SM.

It is therefore of interest to apply the Madala hypothesis to $tth$ results, in order to understand whether it can shed light on the excesses seen in the data.
The Madala boson, if it is produced in association with top quarks, can decay into a Higgs boson, therefore mimicking the signature searched for in the existing experimental $tth$ search channels.

\section*{Modelling top associated Madala production}
\label{sec:model}

\begin{figure}
  \centering
  \subfloat[]{\includegraphics[scale=0.65]{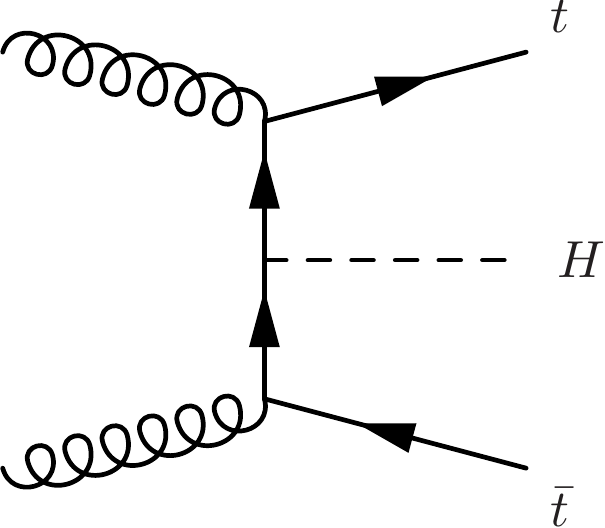}}
  \quad\quad\quad\quad
  \subfloat[]{\includegraphics[scale=0.75]{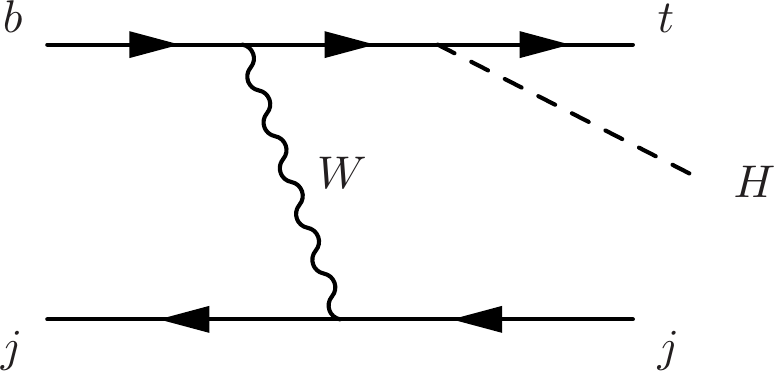}}
  \caption{The leading order (LO) Feynman diagrams for (a) $ttH$ and (b) $tH$ production. Note that the diagram on the right assumes a five-flavour (5F) proton -- that is, $b$-quarks are assumed to be a non-negligible component of the partonic structure of the proton.}
  \label{fig:diagrams}
\end{figure}

Typically in $tth$ searches, signal contributions can come both from Higgs production in association with one or two top quarks, labelled as $th$ and $tth$, respectively.
This is no different for top associated $H$ production.
The dominant Feynman diagrams for top associated $H$ production can be seen in \autoref{fig:diagrams}.

In the SM, it is well known that due to the negative interference between the Yukawa couplings and the Higgs couplings to the weak vector bosons, the production cross section of $th$ is far smaller than that of $tth$.
This is not true for $tH$ production, however.
Following the logic in Ref.~\cite{Farina:2012xp} and the fact that $H$ is assumed to couple weakly to the vector bosons~\cite{vonBuddenbrock:2015ema}, it turns out that
\begin{equation}
  \sigma_{tH}\simeq\sigma_{ttH},
\end{equation}
and so both $tH$ and $ttH$ are non-negligible processes in the Madala hypothesis.
The top Yukawa coupling to the $H$ is assumed to be Higgs-like, and further scaled by the free dimensionless parameter $\beta_g$.
Therefore, one can find the $ttH$ production cross section by first taking the associated value for a heavy Higgs boson from the CERN yellow book~\cite{deFlorian:2016spz}, and then by multiplying it by $\beta_g^2$.
In the Run 1 fit result~\cite{vonBuddenbrock:2015ema}, $\beta_g$ was constrained to be $1.5\pm0.6$.

For this short paper, we assume that $H$ has one dominant decay mode, $H\to Sh$.
Furthermore, $S$ is chosen to be Higgs-like, such that all branching ratios (BRs) are already determined; this choice drastically reduces the number of free parameters in the model.
This choice also enhances the number of leptons one would find as a result of the decay of $S$, since for a Higgs-like particle with a mass near $2m_W$, the BR for $S\to WW$ becomes dominant.
These $W$ bosons in the final state can then decay leptonically to provide a source of multiple lepton production.
In particular, a non-negligible production of two same-sign leptons is possible through the cascade decays, a process that is highly suppressed in the SM.

To simulate $tH$ and $ttH$ production, the hard scatter processes were produced at LO using Monte Carlo (MC) event generation in \textsc{MadGraph}~\cite{Alwall:2014hca}.
This was done using a custom-designed model file using the \textsc{Universal FeynRules Output} (UFO)~\cite{Alloul:2013bka}.
These MC events were passed to \textsc{Pythia 8.2}~\cite{Sjostrand:2014zea} for the resonance decays, parton shower (PS), and hadronisation processes.
Finally, the \textsc{Pythia} output was run through the \textsc{Delphes 3} fast detector simulation~\cite{deFavereau:2013fsa} to account for detector effects.
The output from \textsc{Delphes} was used for the analysis presented in the next section.

\section*{Comparisons with CMS data}

A relatively recent experimental result (at the time of writing) was chosen for the comparison of the Madala hypothesis with data from the LHC.
The chosen experimental result is a search done by the CMS collaboration for a single top quark in association with a Higgs boson~\cite{CMS-PAS-HIG-17-005}.
It should be noted upfront that even though it has been labelled as a ``single top'' search, the event selection (shown in \autoref{tab:selection}) is compatible with both single and double top associated production of both the Higgs and Madala bosons.

The analysis done in the CMS Run 2 single top search is performed by a boosted decision tree (BDT), so a direct comparison with their final results is difficult.
However, the paper did present distributions of three key variables that were made after an event preselection detailed in \autoref{tab:selection}.
These variables are: the largest absolute pseudo-rapidity of any jet in the event, the separation in azimuthal angle between the leading same-sign lepton pair, and the jet multiplicity.

\begin{table}
  \centering
  \caption{Summary of the event selection and categorisation for the CMS Run 2 single top search.
  These criteria closely mimic the selection done in Ref.~\cite{CMS-PAS-HIG-17-005}, since a comparison to data is performed.
  This selection is used on the output $n$-tuple created by the \textsc{Delphes} fast detector simulation.}
  \begin{tabular}{|c|c|}
    \hline
    \multicolumn{2}{|c|}{\textbf{Event selection}} \\
    \hline
    \multicolumn{2}{|c|}{No lepton pair with $m_{\ell\ell} < 12$~GeV} \\
    \multicolumn{2}{|c|}{$N_{b\text{-jets}}\geq1$} \\
    \multicolumn{2}{|c|}{$N_{\text{jets}}\geq1$ (not including $b$-jets)} \\
    \hline
    \multicolumn{2}{|c|}{\textbf{Event categorisation}} \\
    \hline
    \textit{Same-sign 2 lepton} & \textit{Tri-lepton} \\
    \hline
    Exactly 2 same-sign leptons & Exactly 3 leptons \\
    $\ell\ell=e\mu~\text{or}~\mu\mu$ & Leading lepton \pT $>$ 25~GeV \\
    Leading lepton \pT $>$ 25~GeV & Second and third lepton \pT $>$ 15~GeV \\
    Sub-leading lepton \pT $>$ 15~GeV & No lepton pair with $\left|m_{\ell\ell}-m_Z\right|<15$~GeV \\
    \hline
  \end{tabular}
  \label{tab:selection}
\end{table}

To compare the Madala hypothesis with the data presented by the CMS Run 2 single top search, BSM events were generated and run through a CMS detector simulation as detailed in the section above.
Only one mass point was considered for the analysis, since the acceptances into the preselection regions of the chosen analysis were determined to be not significantly sensitive to the change in the mass of $S$.
Rather, the ability of the process to produce a final state with two same-sign leptons is what fundamentally determines the preselection acceptance, and in the proposed mass range for $S$, this does not change significantly since the BRs of $S$ do not change significantly.
The mass of $H$ was set to 270~GeV (in line with the Run 1 fit result~\cite{vonBuddenbrock:2015ema}), while the mass of $S$ was set to 140~GeV, such that the $H\to Sh$ decay could be kept on-shell.

\begin{figure}
  \centering
  \subfloat[]{\includegraphics[width=0.49\textwidth]{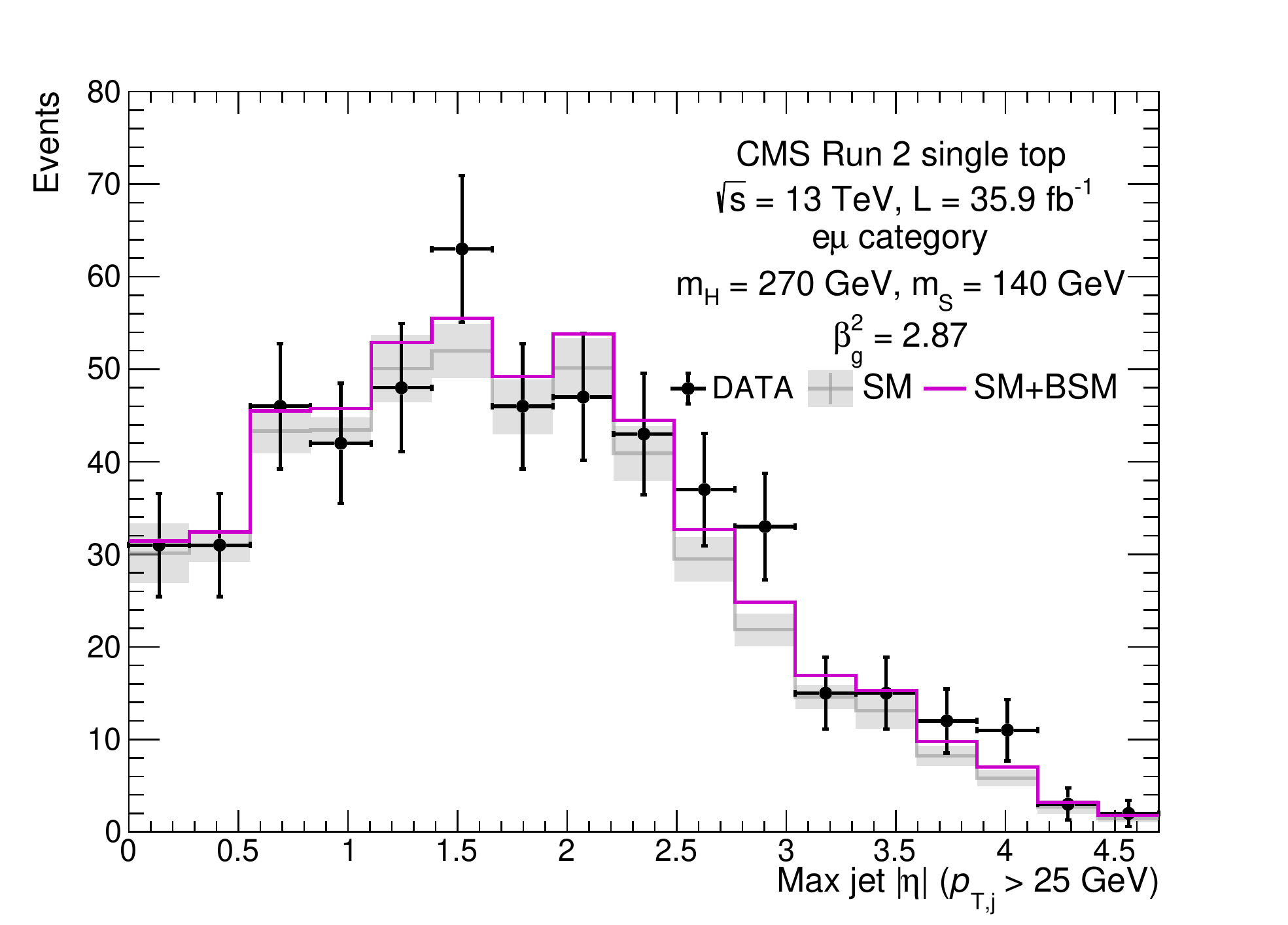}}
  \subfloat[]{\includegraphics[width=0.49\textwidth]{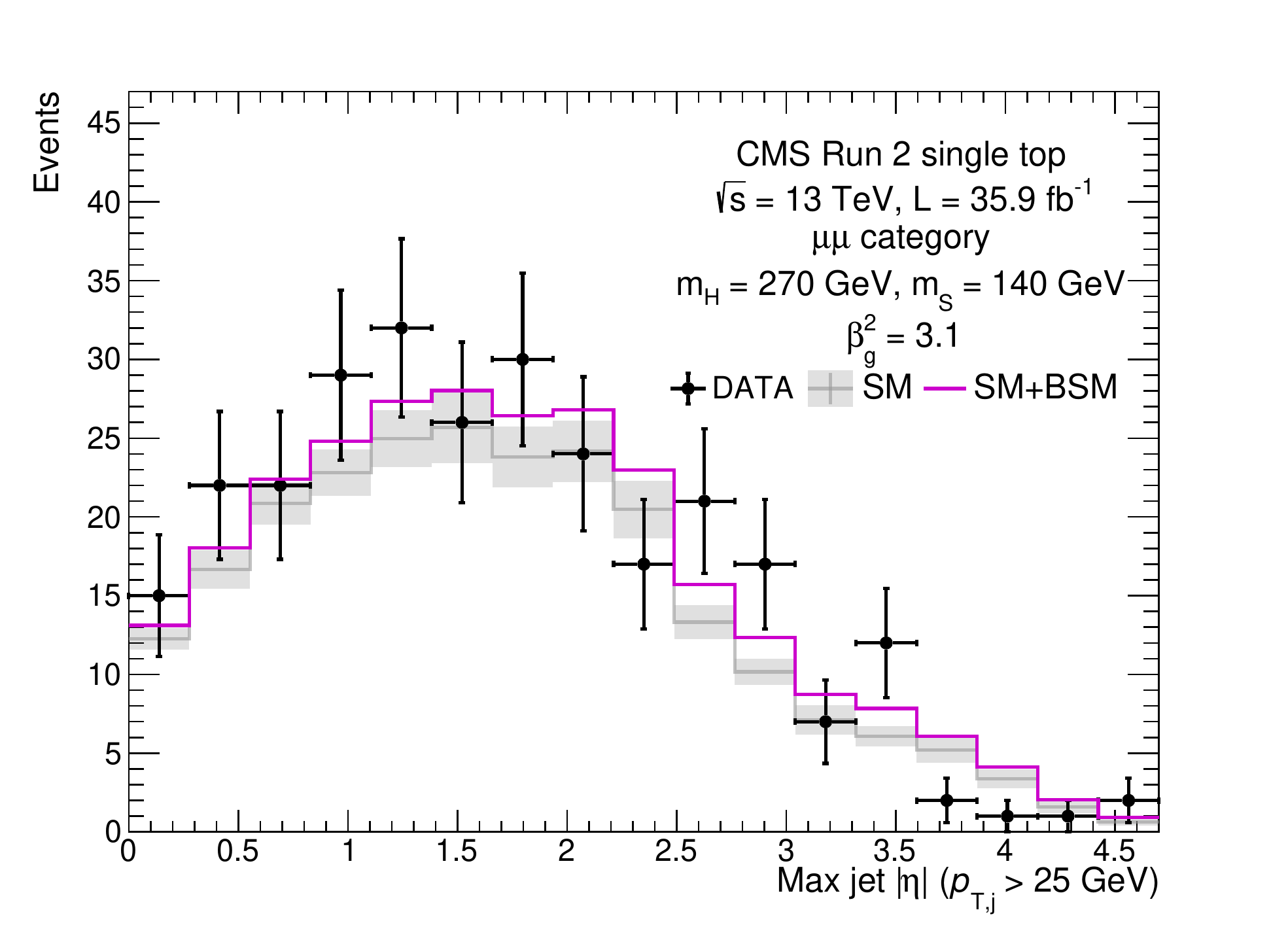}}

  \subfloat[]{\includegraphics[width=0.49\textwidth]{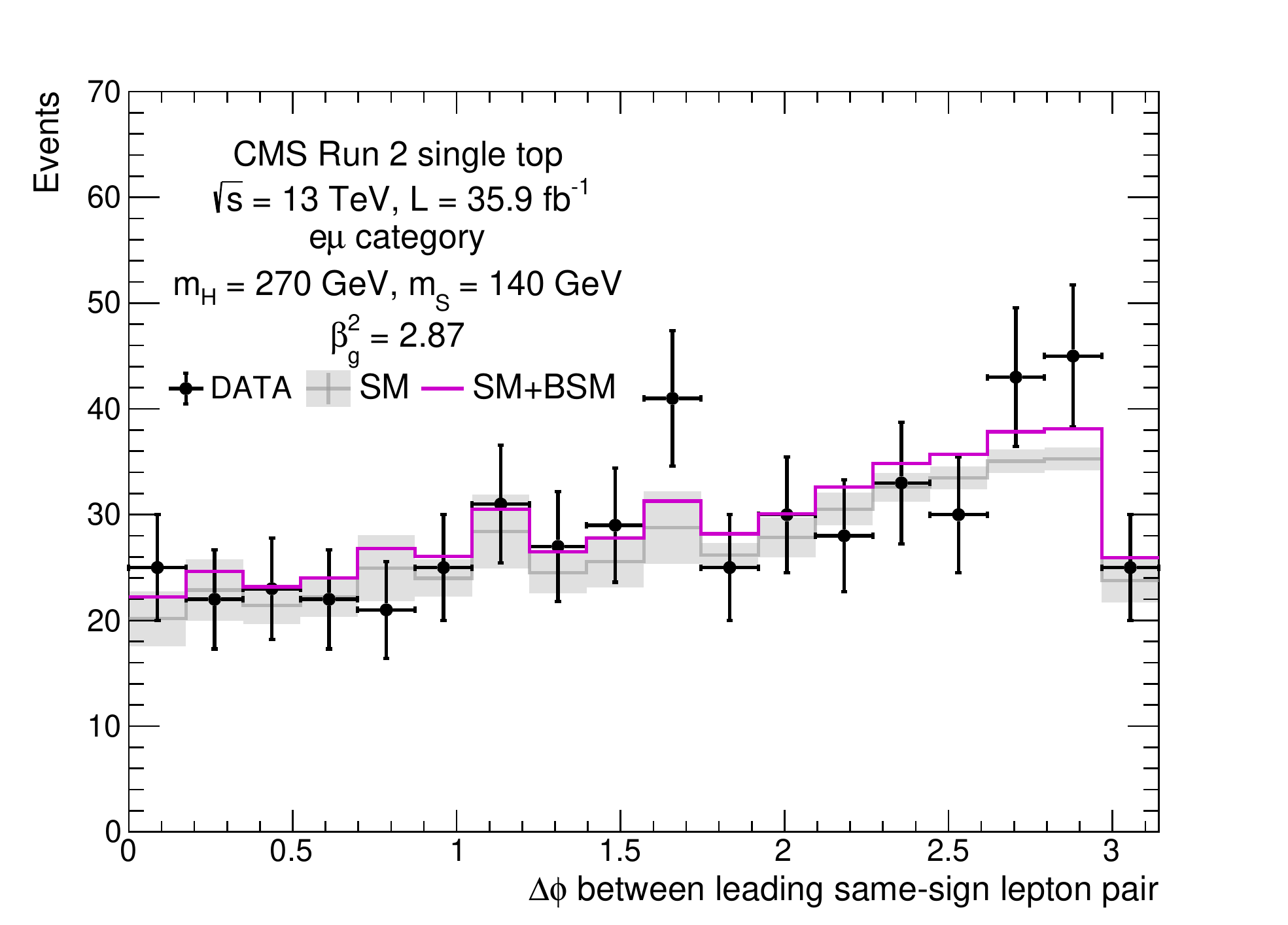}}
  \subfloat[]{\includegraphics[width=0.49\textwidth]{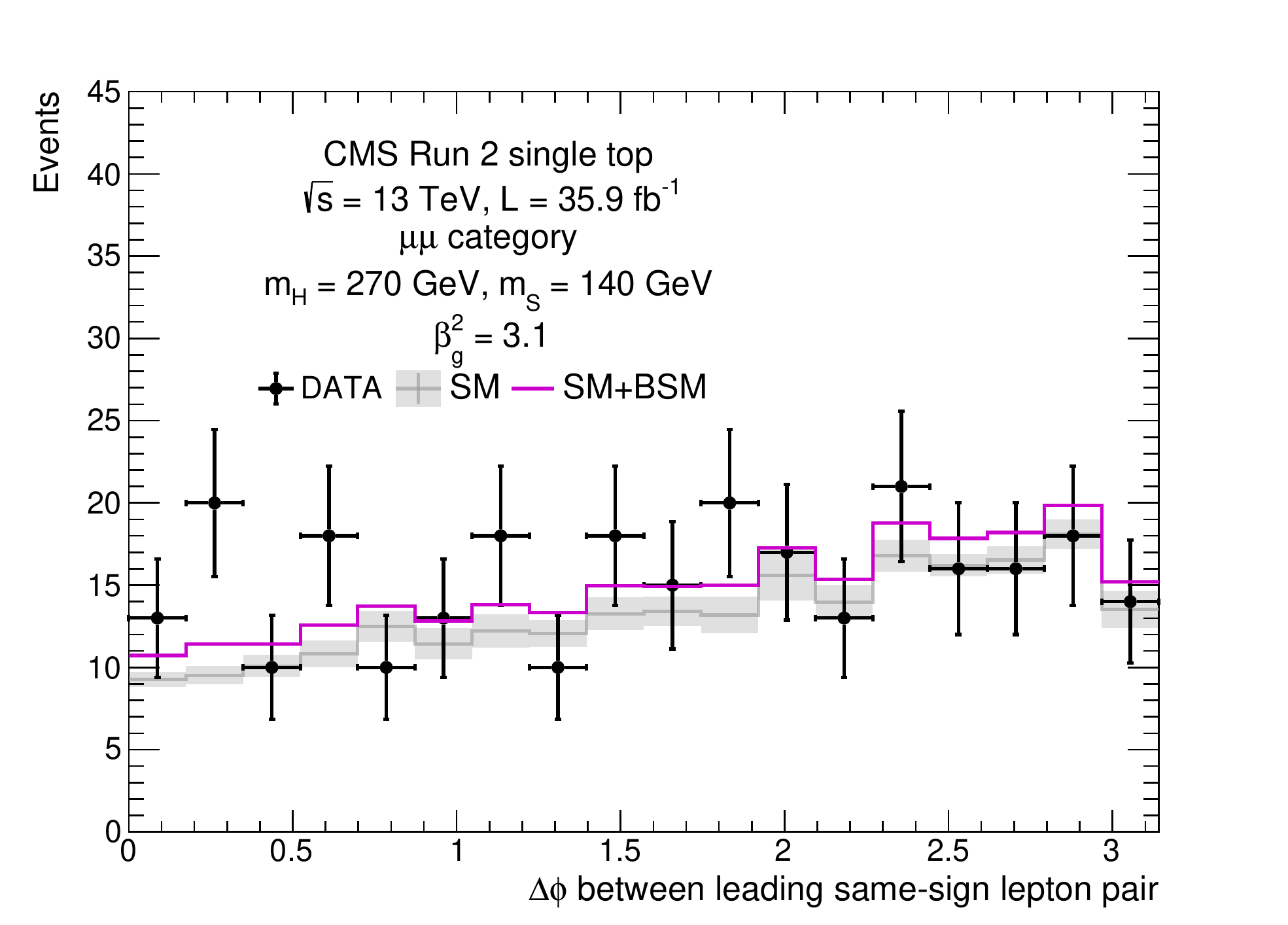}}

  \subfloat[]{\includegraphics[width=0.49\textwidth]{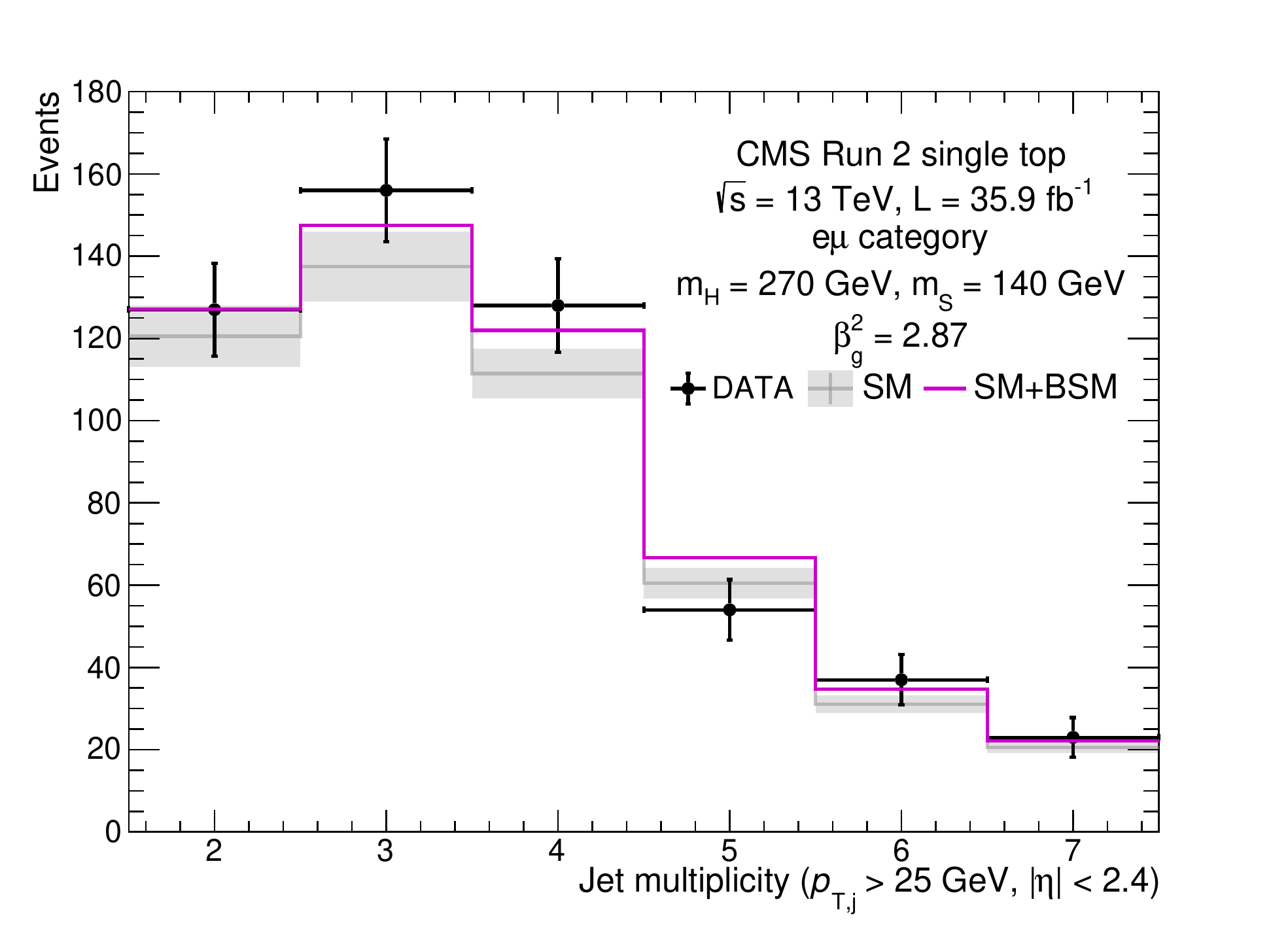}}
  \subfloat[]{\includegraphics[width=0.49\textwidth]{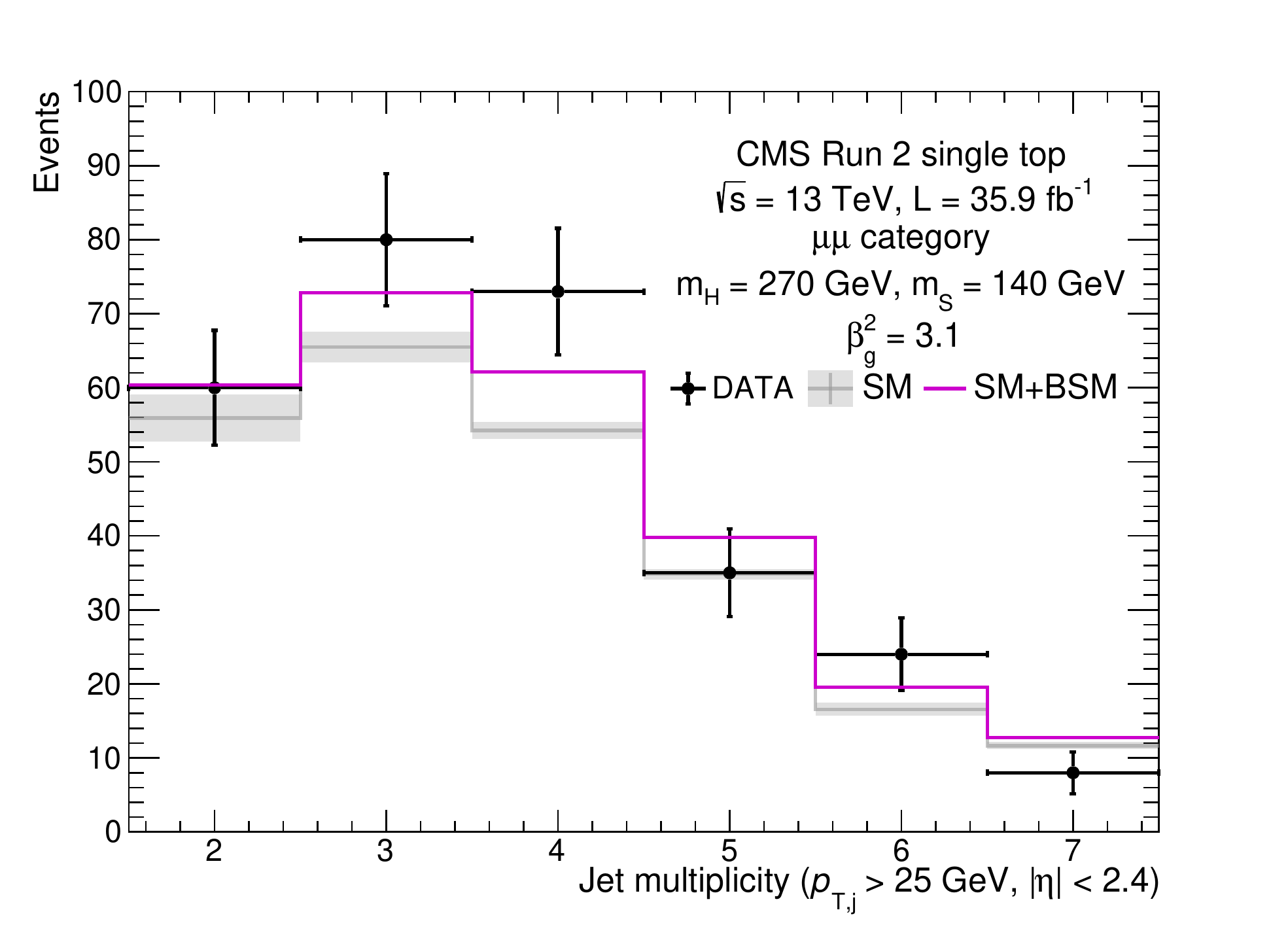}}

  \caption{The plots of (a-b) maximum jet pseudo-rapidity, (c-d) azimuthal separation between the leading same-sign lepton pair and (e-f) jet multiplicity from the CMS Run 2 single top search. These have been separated into the $e\mu$ channel (left) and the $\mu\mu$ channel (right). The tri-lepton channel did not yield a positive BSM signal, so the plots have been omitted. The BSM predictions are scaled by the best fit values of $\beta_g^2$ as described in the text. The mass points considered for the BSM prediction are $m_H=270$~GeV and $m_S=140$~GeV.}
  \label{fig:plots}
\end{figure}

The events were filtered according to the selection and categorisation detailed in \autoref{tab:selection}, and then plotted as a function of the three key variables listed above.
The SM background and its associated uncertainty was read off of the figures in Ref.~\cite{CMS-PAS-HIG-17-005}.
The BSM prediction was scaled to the appropriate cross section from Ref.~\cite{deFlorian:2016spz} multiplied by a best fit value of $\beta_g^2$ and added to the SM prediction.
The results of this can be seen for the $e\mu$ and $\mu\mu$ channels in \autoref{fig:plots}.
It should be noted that the best fit value of $\beta_g^2$ was negative for the tri-lepton channel, and therefore the plots have not been included in this short paper.

The best fit values of $\beta_g^2$ were computed as follows.
A global $\chi^2$ was constructed by adding a $\chi^2$ for each bin and in each observable per channel.
For each bin $i$, the global $\chi^2$ therefore took the form of Pearson's test statistic,
\begin{equation}
  \chi^2=\sum_i\frac{\left(N_i^\text{data}-N_i^\text{SM}-\beta_g^2N_i^\text{BSM}\right)^2}{\left(\Delta N_i^\text{data}\right)^2+\left(\Delta N_i^\text{SM}\right)^2},
  \label{eqn:chisqaure}
\end{equation}
where the $N_i$ factors are the event yields per bin $i$, and the $\Delta N_i$ factors are their associated uncertainties.
The BSM uncertainty is not included since it is dominated by the SM and statistical uncertainty.
The best fit value of $\beta_g^2$ is obtained by minimising \autoref{eqn:chisqaure} while leaving $\beta_g^2$ free.
A $1\sigma$ uncertainty on this best fit value is taken as the envelope around which \autoref{eqn:chisqaure} can vary by one unit away from the mean fit value of $\beta_g^2$.
The results of this process can be seen in \autoref{tab:results}.

\begin{table}
  \renewcommand\arraystretch{1.25}
  \caption{The best fit values for $\beta_g^2$ for each channel in the CMS Run 2 single top search. The combined result was obtained by summing \autoref{eqn:chisqaure} over all channels. The Run 1 fit result from Ref.~\cite{vonBuddenbrock:2015ema} has been listed to show the compatibility between the new and old results.}
  \centering
  \begin{tabular}{|c|r|}
    \hline
    \textbf{Channel} & \textbf{Best fit $\beta_g^2$} \\
    \hline
    $e\mu$ & $3.10\pm1.02$ \\
    $\mu\mu$ & $2.87\pm1.04$ \\
    Tri-lepton & $-0.93\pm0.92$ \\
    \hline
    \textbf{Combined} & $1.48\pm0.57$ \\
    \hline
    Run 1 fit result & $2.25\pm1.80$ \\
    \hline
  \end{tabular}
  \label{tab:results}
\end{table}

\section*{Discussion}

As can be seen in \autoref{fig:plots}, the BSM prediction does a relatively good job of explaining the CMS data in the $e\mu$ and $\mu\mu$ channels -- this is possibly best seen in the jet multiplicity distributions.
As noted before, the fit to the tri-lepton data did not yield a positive BSM signal.
Note that a negative value of $\beta_g^2$ is not physical, however the fit value is still included in the results due to the fact that the whole dataset for the CMS search is statistically limited.
In addition to this, the further categorisation of the data into the three channels further limits the statistical reach of the analysis.
It is therefore far more sensible to treat the combined fit value as a result with statistical power, while the fits to the individual categories behave more like statistical fluctuations around the combined best fit value.
Whether or not the poor fit to the tri-lepton channel yields any information about the nature of the model's ability to accurately describe the data in terms of lepton multiplicity is unclear, due to the statistical limitations of the dataset.
The combination of all channels shown in \autoref{tab:results} is still dominated by the di-lepton channels, of which the best fit values of $\beta_g^2$ have relative uncertainties of $\sim35$~\%, whereas for the tri-lepton channel it is $\sim100$~\%.
The combined best fit value of $\beta_g^2$ equates to around a $2.6\sigma$ deviation from the SM.

As mentioned in the introduction, a combination of leptonic $tth$ searches can be quantified by the signal strength parameter $\mu_{tth}=1.92\pm0.38$ as calculated in Ref.~\cite{vonBuddenbrock:2017bqf}.
Using the combined best fit value of $\beta_g^2$ in \autoref{tab:results}, the corresponding value of $\mu_{tth}$ for the analysis done in this short paper is compatible with this, and is equal to $1.32\pm0.51$.
Combining the results of this short paper with the global combination done in Ref.~\cite{vonBuddenbrock:2017bqf}, a significance of $3.5\sigma$ in excess of the SM can be computed.

The CMS Run 2 single top search has corroborated previous results that have been calculated using the Madala hypothesis.
It is still, however, only one analysis that can be analysed from the Run 2 dataset.
It is imperative, then, to continue constraining the model with other data in leptonic $tth$ searches from both CMS and ATLAS.
More importantly, it is prudent to compare the predictions of the Madala hypothesis with differential distributions, rather than single measurements of signal strength, since distributions provide a more statistically rich picture of potential deviations from the SM.
It should be noted also that the simplification of assuming a Higgs-like $S$ particle could also be generalised to more complicated and phenomenologically rich models.
However, so far the compatibility of the results in this paper with previous results from the Madala hypothesis is an encouraging outcome of the study, which will lead to further work done on constraining the properties of the Madala boson.

\newcommand{\newblock}{}
\bibliographystyle{iopart-num}
\bibliography{ref.bib}

\providecommand{\newblock}{}
\begin{thebibliography}{10}
\expandafter\ifx\csname url\endcsname\relax
  \def\url#1{{\tt #1}}\fi
\expandafter\ifx\csname urlprefix\endcsname\relax\def\urlprefix{URL }\fi
\providecommand{\eprint}[2][]{\url{#2}}

\bibitem{Aad:2012tfa}
{ATLAS Collaboration} 2012 {\em Phys. Lett.\/} {\bf B716} 1--29
  (arXiv:\eprint{1207.7214})

\bibitem{Chatrchyan:2012xdj}
{CMS Collaboration} 2012 {\em Phys. Lett.\/} {\bf B716} 30--61
  (arXiv:\eprint{1207.7235})

\bibitem{Csaki:2018muy}
Csáki C, Lombardo S and Telem O 2018 {TASI Lectures on Non-supersymmetric BSM
  Models} Tech. rep.

\bibitem{vonBuddenbrock:2016rmr}
von Buddenbrock S, Chakrabarty N, Cornell A~S, Kar D, Kumar M, Mandal T,
  Mellado B, Mukhopadhyaya B, Reed R~G and Ruan X 2016 {\em Eur. Phys. J.\/}
  {\bf C76} 580 (arXiv:\eprint{1606.01674})

\bibitem{vonBuddenbrock:2015ema}
von Buddenbrock S, Chakrabarty N, Cornell A~S, Kar D, Kumar M, Mandal T,
  Mellado B, Mukhopadhyaya B and Reed R~G 2015  (arXiv:\eprint{1506.00612})

\bibitem{vonBuddenbrock:2017bqf}
von Buddenbrock S 2017 {\em Proceedings of KRUGER2016\/}
  (arXiv:\eprint{1706.02477})

\bibitem{Farina:2012xp}
Farina M, Grojean C, Maltoni F, Salvioni E and Thamm A 2013 {\em JHEP\/} {\bf
  05} 022 (arXiv:\eprint{1211.3736})

\bibitem{deFlorian:2016spz}
de~Florian D {\em et~al.\/} (LHC Higgs Cross Section Working Group
  Collaboration) 2016  (arXiv:\eprint{1610.07922})

\bibitem{Alwall:2014hca}
Alwall J, Frederix R, Frixione S, Hirschi V, Maltoni F, Mattelaer O, Shao H~S,
  Stelzer T, Torrielli P and Zaro M 2014 {\em JHEP\/} {\bf 07} 079
  (arXiv:\eprint{1405.0301})

\bibitem{Alloul:2013bka}
Alloul A, Christensen N~D, Degrande C, Duhr C and Fuks B 2014 {\em Comput.
  Phys. Commun.\/} {\bf 185} 2250--2300 (arXiv:\eprint{1310.1921})

\bibitem{Sjostrand:2014zea}
Sj{\"o}strand T, Ask S, Christiansen J~R, Corke R, Desai N, Ilten P, Mrenna S,
  Prestel S, Rasmussen C~O and Skands P~Z 2015 {\em Comput. Phys. Commun.\/}
  {\bf 191} 159--177 (arXiv:\eprint{1410.3012})

\bibitem{deFavereau:2013fsa}
de~Favereau J, Delaere C, Demin P, Giammanco A, Lema{\^\i}tre V, Mertens A and
  Selvaggi M (DELPHES 3 Collaboration) 2014 {\em JHEP\/} {\bf 02} 057
  (arXiv:\eprint{1307.6346})

\bibitem{CMS-PAS-HIG-17-005}
{CMS Collaboration} 2017 {Search for production of a Higgs boson and a single
  top quark in multilepton final states in proton collisions at
  $\sqrt{s}=13~\mathrm{TeV}$} Tech. Rep. CMS-PAS-HIG-17-005 CERN Geneva
  \urlprefix\url{http://cds.cern.ch/record/2264553}

\end{thebibliography}

\end{document}